\documentclass[conference]{IEEEtran}
\IEEEoverridecommandlockouts
\usepackage{cite}
\usepackage{amsmath,amssymb,amsfonts}
\usepackage{algorithmicx}
\usepackage{graphicx}
\usepackage{textcomp}
\usepackage{xcolor}
\usepackage{listings}
\usepackage{amssymb}
\usepackage{pifont}
\usepackage{tabularx}
\usepackage{xurl}
\usepackage{algorithm}
\usepackage{algpseudocode}

\usepackage{amsmath, amssymb}
\usepackage{tikz}
\usepackage{amsmath}
\usepackage{filecontents}
\usepackage{xcolor}
\usepackage{pifont}
\newcommand{\cmark}{\ding{51}}%
\newcommand{\xmark}{\ding{55}}%
\lstdefinestyle{mystyle}{
    language=Python,
    commentstyle=\color{olive},
    keywordstyle=\color{blue},
    numberstyle=\tiny\color{black},
    numbersep=5pt,
    breaklines=true,
    showstringspaces=false,
    basicstyle=\ttfamily\scriptsize,
    frame=single,
    stringstyle=\color{black},
    morekeywords={join, filter_sql, with_column, groupby, agg, collect},
}

\def\BibTeX{{\rm B\kern-.05em{\sc i\kern-.025em b}\kern-.08em
    T\kern-.1667em\lower.7ex\hbox{E}\kern-.125emX}}
\begin{document}

\title{Efficient Fault Tolerance for Pipelined Query Engines via Write-ahead Lineage}

\author{\IEEEauthorblockN{Ziheng Wang}
\IEEEauthorblockA{
\textit{Stanford University}\\
Stanford, CA, U.S.A \\
zihengw@stanford.edu}
\and
\IEEEauthorblockN{Alex Aiken}
\IEEEauthorblockA{
\textit{Stanford University}\\
Stanford, CA, U.S.A \\
aaiken@stanford.edu}

}

\maketitle

\begin{abstract}
Modern distributed pipelined query engines either do not support intra-query fault tolerance or employ high-overhead approaches such as persisting intermediate outputs or checkpointing state. In this work, we present write-ahead lineage, a novel fault recovery technique that combines Spark's lineage-based replay and write-ahead logging. Unlike Spark, where the lineage is determined before query execution, write-ahead lineage persistently logs lineage at runtime to support dynamic task dependencies in pipelined query engines. Since only KB-sized lineages are persisted instead of MB-sized intermediate outputs, the normal execution overhead is minimal compared to spooling or checkpointing based approaches. To ensure fast fault recovery times, tasks only consume intermediate outputs with persisted lineage, preventing global rollbacks upon failure. In addition, lost tasks from different stages can be recovered in a pipelined parallel manner. We implement write-ahead lineage in a distributed pipelined query engine called Quokka. We show that Quokka is around 2x faster than SparkSQL on the TPC-H benchmark with similar fault recovery performance.
\end{abstract}

\section{\textbf{Introduction}}

In the past decade, the ``data lake'' has emerged as a popular paradigm for data management in the cloud. In a data lake, data is stored directly in object storage on a cloud service such as AWS S3, in open formats like CSV and Parquet. Distributed query engines like SparkSQL and Trino read this data and shuffle it across the network to execute SQL queries \cite{armbrust2015spark, melnik2010dremel, sethi2019presto}.

The first generation of distributed query engines, such as MapReduce and SparkSQL, relied on a bulk synchronous execution model similar to MPI \cite{dean2008mapreduce, zaharia2016apache}. While these frameworks offer flexibility and efficient lineage-based fault tolerance \cite{zaharia2016apache}, they were originally designed for iterative tasks such as PageRank or machine learning training. When repurposed for data analytics tasks such as join-heavy SQL queries, their stagewise execution model leaves performance on the table by foregoing parallelism across stages commonly exploited in pipelined query engines \cite{shen2020magnet}.

The second generation of distributed query engines were purpose-built for SQL, and often employ a pipelined architecture where multiple stages can execute concurrently \cite{sethi2019presto, dageville2016snowflake}. In some of these systems, the dependencies between tasks in different stages are statically determined \cite{sethi2019presto}. In other more recent systems, such dependencies are dynamically determined to take advantage of work stealing and cache efficiencies \cite{armenatzoglou2022amazon, leis2014morsel}. While these engines typically drastically outperform SparkSQL, they either do not support intra-query fault tolerance\footnote{While most cloud query engines are fault tolerant to data loss, this paper focuses on intra-query fault tolerance, which means the query engine can reuse intermediate results to recover faster than restarting the entire query after a machine failure.} and re-execute failed queries from the beginning \cite{dageville2016snowflake, armenatzoglou2022amazon}, or rely on high-overhead approaches such as durably persisting shuffle partitions between stages \cite{trino-ft}.\footnote{It is important to distinguish this \textit{overhead}, paid during normal operation to support fault recovery, from \textit{recovery performance} after an actual failure has occurred.} 

Fault-tolerant streaming systems, such as Kafka Streams, Apache Flink or StreamScope \cite{kreps2011kafka, carbone2015apache, lin2016streamscope}, are also based on a pipelined architecture and may feature an SQL interface like FlinkSQL. However, their fault tolerance strategies are typically optimized for continuous processing of small fresh inputs, instead of bulk shuffles of large batches of data. As a result, their checkpointing-based fault tolerance strategies impose high overheads during normal execution for batch SQL processing.


In this paper, we adapt lineage-based fault tolerance to pipelined query engines with dynamic task dependencies. Unlike Spark, where lineage is statically determined and fault recovery is data parallel, we consistently log the lineage after it has been dynamically determined at runtime and use pipelined parallel fault recovery  \cite{zaharia2016apache, zaharia2012resilient}.   We term this novel fault tolerance strategy \textbf{write-ahead lineage}. We implement this strategy in an open source distributed query engine called Quokka.

Write-ahead lineage allows Quokka to only persist KB-sized lineage information in normal operation instead of MB-sized shuffle partitions or GB-sized state checkpoints. Restricting tasks to consume only intermediate outputs whose lineage has been persisted limits fault recovery to only tasks scheduled on failed workers, avoiding the expensive globally coordinated rollbacks in checkpointing-based systems. Pipelined parallel recovery further speeds up the fault recovery process, allowing Quokka to achieve similar recovery performance as Spark. 

To the best of our knowledge, Quokka is the only distributed query engine with a dynamic pipelined execution model that supports low-overhead intra-query fault tolerance. We believe write-ahead lineage can also be easily implemented in other pipelined query engines. This paper makes two key contributions: 

\begin{itemize}
  \setlength\itemsep{0em}
    \item We describe \textbf{write-ahead lineage}, a novel fault tolerance strategy for pipelined query engines with dynamic task dependencies based on consistent lineage logging and parallel replay.
    \item We implement a distributed query engine, Quokka. We demonstrate write-ahead lineage allows Quokka to support fault tolerance with low execution overhead and fast fault recovery. On TPC-H, Quokka outperforms SparkSQL by around 2x on up to 32 nodes in normal execution, with competitive fault recovery performance.
\end{itemize}

\section{\textbf{Motivations From Related Work}}


There are decades of related work on pipeline query engines \cite{sethi2019presto, leis2014morsel, shaikhha2018push, melnik2010dremel, neumann2011efficiently, harizopoulos2005qpipe, boncz2005monetdb} and fault tolerance in dataflow systems \cite{murray2013naiad, akidau2013millwheel, kreps2011kafka, lin2016streamscope, carbone2015apache}. First, let us define the scope of our paper by defining what is \textit{dynamic pipelined execution} in the context of a distributed query engine.

\subsection{\textbf{Dynamic Pipelined Execution}}

Modern distributed SQL query engines such as Snowflake, AWS RedShift and Trino all adopt a streaming architecture, where data is ``pushed'' through a pipeline of multiple stages \cite{dageville2016snowflake, sethi2019presto, armenatzoglou2022amazon}. Such pipelined execution typically results in higher available parallelism, resource utilization and cache efficiency than SparkSQL's one-stage-at-a-time approach. 

\begin{figure}
    \centering
    \includegraphics[width=\linewidth,keepaspectratio]{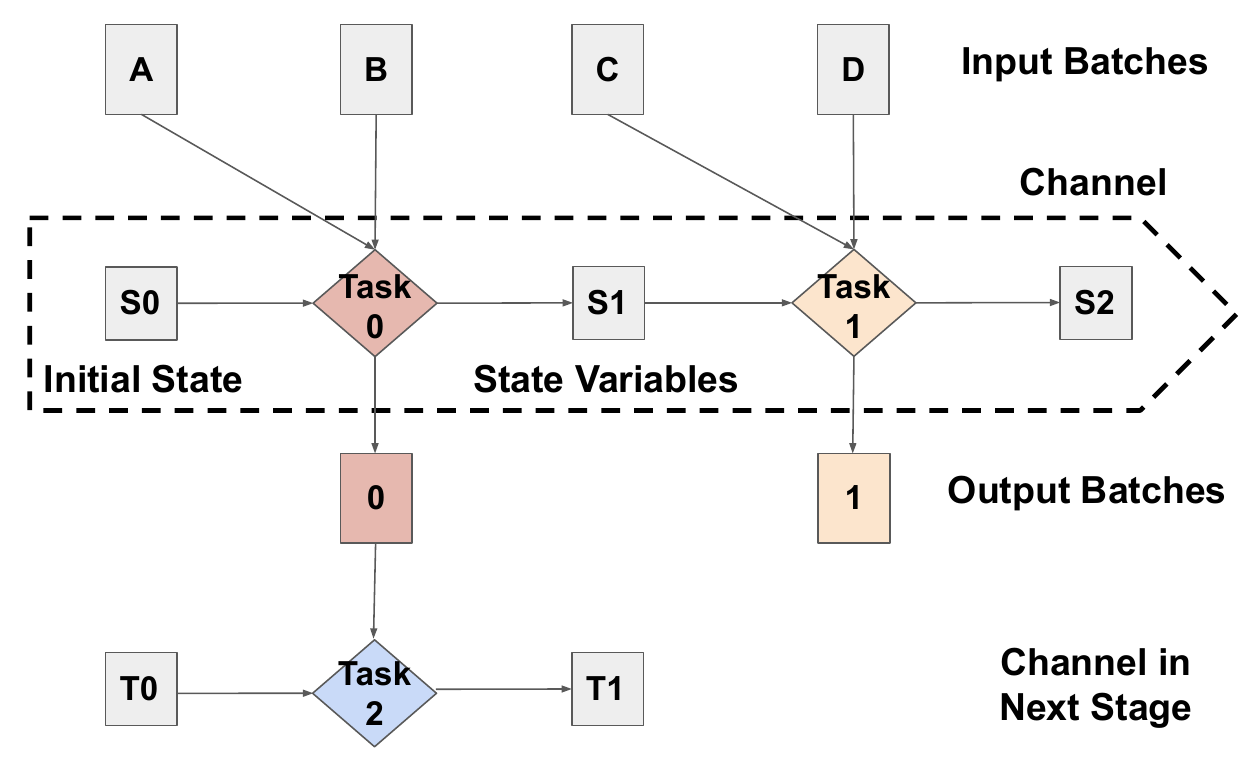}
    \caption{
        Pipelined Engine: An input stream of batches is processed by tasks to create an output stream, producing intermediate state variables along the way. The number and size of batches processed by each task can be dynamically determined at runtime.
        \label{fig:rds}
        \vspace{-0.5em}
    }
\end{figure}

In a pipelined query execution engine, outputs produced by tasks in one stage can be immediately consumed by tasks in the next stage. Tasks associated with some operators, such as joins and aggregations, might have an associated \textit{state variable} as illustrated in Figure \ref{fig:rds}. Subsequent tasks in a stage need to account for the state variable generated by previous tasks. For example, if the stage is building a hash table for a build-probe join, A, B, C and D could be batches from the build side, while the state variable corresponds to the hash table. S0 would be empty while S2 would be the complete hash table.

In Figure \ref{fig:rds}, task 1 and task 2 can be scheduled concurrently, exposing pipeline parallelism across different stages. Note this execution model does not preclude data parallelism as in SparkSQL. Each stage can still be executed by different parallel \textit{channels}, where each channel consists of a sequence of tasks. Each channel operates on a different portion of the data, e.g. separate hash partitions of the join key. Each channel can be scheduled on a physical machine, and may have its own associated state variable. Tasks in a channel could depend on tasks from all the channels in the previous stage.

In a pipelined query engine, the task dependencies could be \textit{static} or \textit{dynamic}. In a static pipelined query engine, the number of input batches, the number of tasks and channels in each stage and their dependencies are statically determined before query execution \cite{sethi2019presto}. More recent distributed query engines such as the newest Amazon Redshift version typically determine these task dependencies dynamically at runtime \cite{armenatzoglou2022amazon}, inspired by recent designs such as work-stealing and morsel-driven parallelism \cite{leis2014morsel}.



In a dynamic pipelined query engine, assigning task dependencies at runtime achieves similar work-stealing and cache-efficiency benefits as in single-node multi-core query engines that pioneered this design \cite{leis2014morsel}. For example, allowing task 0 in Figure \ref{fig:rds} to be launched as soon as any of input batches A, B, C or D appear minimizes idle resources. Tasks can also decide how many input batches to take as input based on their own performance characteristics to improve execution efficiency \cite{armenatzoglou2022amazon}. 

\begin{table*}
    \centering
    \caption{Fault tolerance design choices in data processing systems.}
    \begin{tabular}{|l|l|l|l|l|l|l|l|l|l|}
    \hline
          & \textbf{Trino} & \textbf{SparkSQL} & \textbf{Kafka Streams} & \textbf{Flink} & \textbf{StreamScope} & \textbf{Quokka}  \\ \hline
       \textbf{Description} & Pipelined SQL & Stagewise SQL & Dataflow & Dataflow & Dataflow & Pipelined SQL \\ \hline
       \textbf{Spooling} & \cmark & \xmark & \cmark & \xmark & \xmark & \xmark \\ \hline
        \textbf{State Checkpoint} & \xmark & \xmark & \cmark & \cmark & \cmark & \xmark  \\ \hline
        \textbf{Lineage} & \cmark & \cmark & \cmark & \xmark & \cmark & \cmark\\ \hline
    \end{tabular}
    \label{tab:fault}
    
\end{table*} 

\subsection{\textbf{Efficient Fault Tolerance}} \label{ft-all}

The principal question we aim to answer in this paper is: \textbf{how to support fault tolerance for dynamic pipelined query engines with low overhead and fast recovery}? To motivate our novel write-ahead lineage approach, we explore existing approaches and describe their weaknesses here.

Most distributed pipelined query engines do not support intra-query fault tolerance, instead relying on query-retries when a cloud worker instance fails \cite{dageville2016snowflake, armenatzoglou2022amazon}. Only Trino recently added support for intra-query fault tolerance based on HDFS spooling of shuffle partitions \cite{trino-ft}. 

While intra-query fault tolerance for pipelined query engines is still relatively new, they are architecturally similar to stream processing systems based on the dataflow model \cite{murray2013naiad}, differing mainly in the granularity of the data that is passed between different tasks. Fault tolerance for such streaming systems has been thoroughly studied in the past decade, with strategies coalescing around three core techniques, similar to the taxonomy proposed by Falkirk Wheel \cite{gog2021falkirk}:

\begin{itemize}
  \setlength\itemsep{0em}
    \item \textbf{Lineage}: Lineage refers to dependencies between data partitions. It could be persisted to be consulted upon failure to facilitate recovery.
    \item \textbf{Spooling vs Upstream Backup}: Shuffle data partitions, like the colored boxes 0 and 1 in Figure \ref{fig:rds}, maybe stored reliably (\textit{spooling}), unreliably, e.g. on local disk of the producer (\textit{upstream backup}), or not at all. 
    \item \textbf{Checkpointing}: state variables, like S1 and S3 in Figure \ref{fig:rds}, can be persisted.
\end{itemize}

There are other distributed futures and workflow systems like Ray and Exoflow which may resemble pipelined query engines. However, their fault tolerance mechanisms typically focus on handling tasks that might produce different outputs from the same input \cite{moritz2018ray, zhuang2023exoflow}. While they still rely on these three core techniques, their design goals are ifferent from pipelined query engines, whose tasks are all deterministic. As a result, we don't cover these systems in detail here.

Table 1 summarizes the fault tolerant systems we compare in this section, along with which of the three strategies they employ. Trino and SparkSQL are distributed query engines, while Kafka Streams, Flink and StreamScope are stream processing engines. Now we explain how each of the three strategies are applied in these five systems.

\subsubsection{\textbf{Lineage}}

Lineage, the dependencies between data partitions, is a key tool in enabling fault tolerance. Lineage allows a task to guarantee a fixed output by remembering what inputs were used, assuming the task is deterministic.

Of all the systems, Flink is the only one which does not track lineage \cite{carbone2015apache}. Upon fault recovery, failed tasks that need to be relaunched can use different inputs the second time around \cite{flink-ft}. Critically, this decision means tasks could emit different outputs and workers who previously consumed the outputs of the failed task also have to be rewound, which typically results in expensive coordinated global rollbacks of all the channels in the entire system.

All the other systems listed in Table 1 determine lineage statically. Trino and Spark determine task dependencies before the query graph is executed \cite{trino-ft, zaharia2016apache}, while real-time systems such as Kafka Streams and StreamScope rely on the unique event time associated with each record to impose a deterministic execution order among inputs \cite{kreps2011kafka,lin2016streamscope}. In either case, this lineage is assumed to be available after a worker fails. 

\begin{figure}
    \centering
    \includegraphics[width=\linewidth,keepaspectratio]{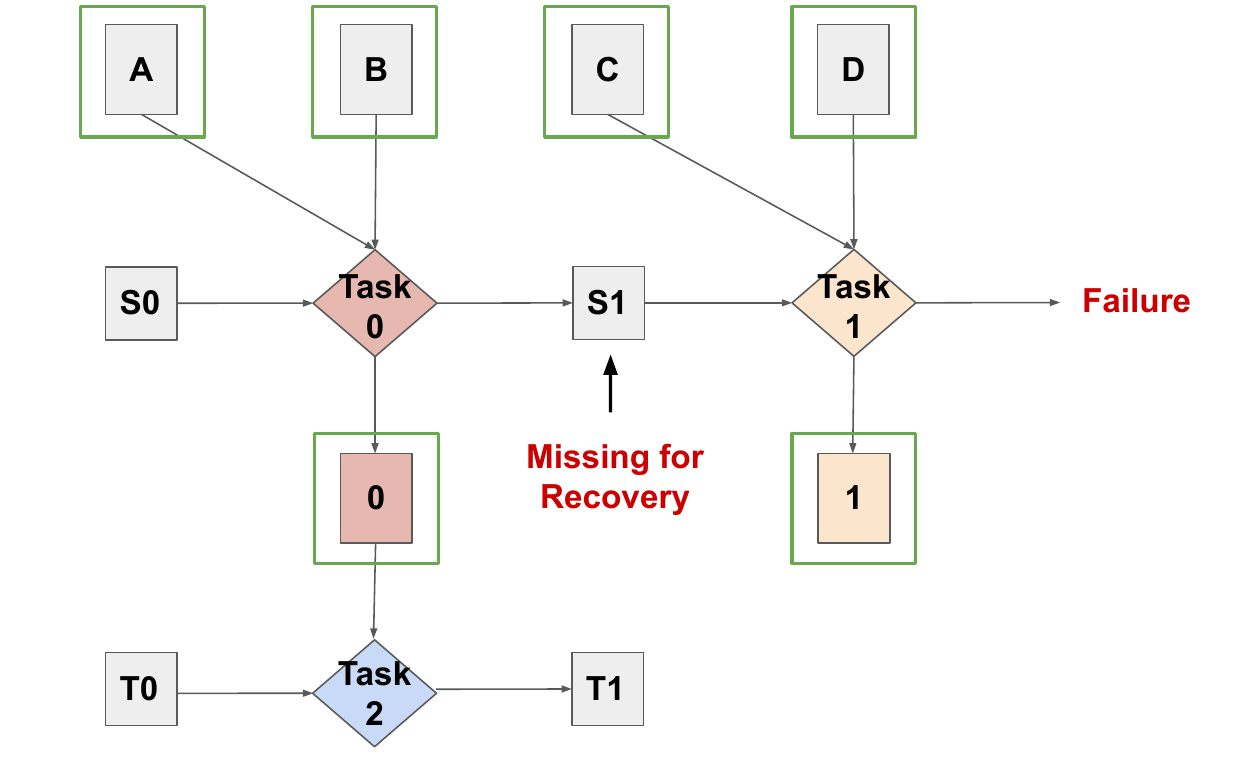}
    \caption{
        Spooling: persisted data partitions are marked with green boxes. We assume the channel with tasks 0 and 1 has failed. Since task 1 depends on state variable S1, which was not persisted, the whole channel has to restart. 
        \label{fig:spool}
        \vspace{-0.5em}
    }
\end{figure}

\subsubsection{\textbf{Upstream Backup / Spooling}}

If we decide to track lineage, making use of it upon fault recovery requires some way of replaying a task's inputs using the lineage, which typically requires storing intermediate data partitions. MapReduce pioneered this approach by persisting reducer outputs in GFS to provide fault tolerance boundaries between different stages \cite{dean2008mapreduce}. Trino stores intermediate data partitions durably in HDFS or an object storage like S3, while Kafka Streams persists them in Kafka topics \cite{kreps2011kafka,trino-ft}. 

However, persisting data partitions, a.k.a spooling, can introduce severe overheads in normal operation, especially in batch analytics. Persisting a data partition in a distributed cluster where workers might fail means either replicating the data partition across the cluster, e.g. Kafka topic or HDFS, or writing the data partition to a blob storage, e.g. Amazon S3. In either case, this operation consumes precious network I/O resources that could be used for the task itself. 


In contrast, Spark relies on unreliable upstream backup of data partitions to local disk of the producer, which is assumed to be lost upon worker failure. Instance-attached NVMe drives have become ubiquitous on public cloud providers, making writing to local disk very efficient compared to persistent writes to HDFS or S3, though the contents of such drives are lost upon worker failure. Avoiding spooling is a key reason why Spark is faster than MapReduce \cite{zaharia2016apache}.

A bigger problem for spooling in a pipelined engine is that it might not save that much work upon failure. The core benefit of spooling is the localization of task retries. In a system relying on upstream backup, if the input to a task that must be retried is also lost, then the task that generated that input must also be retried. Spooling avoids this problem, but only if all of the inputs for the failed task have been persisted.

Unfortunately, in pipelined query engines, as depicted in Figure \ref{fig:rds}, tasks also depend on the channel's state variable. In Figure \ref{fig:spool}, we illustrate what data partitions are persisted in a typical spooling strategy. If the channel experiences a failure after executing task 1, it restarts execution from the beginning state S0 and re-executes task 0, even though all of task 1's input data partitions have been persisted. This is because re-executing task 1 also relies on the state variable S1, which has not be persisted. Streaming engines that perform spooling also commonly ``checkpoint'' these state variables.

\subsubsection{\textbf{Checkpointing}}

We could prevent restarting the failed channel from scratch if we periodically persist the lost state variables. Checkpointing executor state periodically is a popular fault tolerance strategy in real-time streaming systems such as Apache Flink, Kafka Streams and StreamScope. Since jobs in these systems could be continuously operating for days, restarting a channel entirely might cause unacceptable fault recovery performance. Checkpointing also allows the system to garbage collect spooled data partitions, since a data partition does not have to be replayed if its effect has been persisted into a state checkpoint.

However, checkpointing can be even more expensive than spooling for pipelined query engines optimized for SQL queries on large batches of data. Streaming systems typically go to great lengths to ensure that the state of an operator is bounded in size. SQL query engines have no such requirements: consider an operator that builds a hash table for joins. The size of the hash table grows linearly with the number of unique keys it sees. Assuming new keys arrive at a constant rate, naive periodic checkpointing will incur $O(N^2)$ storage complexity where $N$ is the number of unique keys, which can be very large.

Incremental checkpointing could be employed to checkpoint only differences between adjacent checkpoints. While we can easily devise incremental checkpointing strategies for individual stateful operators, efficient generic incremental checkpointing strategies for arbitrary data structures is still an open research problem. Current approaches include persisting a ``changelog'' of the state as in Kafka Streams or leveraging RocksDB's compaction mechanism as in Apache Flink \cite{kreps2011kafka, carbone2015apache}. Both impose heavy constraints on the underlying data structures of the state variable, which is not desirable for a high performance query engine.


\subsubsection{\textbf{Conclusion}}

Our discussion suggests that of the three core fault tolerance techniques for dataflow systems, lineage is the most beneficial for pipelined query engines. 
Spooling incurs significant overheads in normal execution, but is not particularly useful in reducing fault recovery times in the absence of checkpointing. Checkpointing is even more expensive than spooling in terms of overhead and might require undesirable constraints on the state variables.

In the context of pipelined query execution, fault tolerance's first priority should be low overhead. If we cannot achieve low overhead in normal pipelined execution, we are better off running without fault tolerance and retrying queries that fail or using blocking alternatives such as SparkSQL. This analysis naturally leads us to use only lineage instead of spooling and checkpointing.

\section{\textbf{Write-ahead Lineage}}\label{ft-theory}

The core challenge we address is tracking lineage in a pipelined system with dynamic task dependencies and facilitating efficient recovery. While deterministic or static lineage information (as in Spark) can easily be recorded before query execution, dynamically generated lineage is trickier to handle. Unlike Spark, Quokka adopts the write-ahead logging common in other database applications, similar to recent works targeting distributed futures system \cite{wang2019lineage, wang2021ownership}. The lineage of an object must be persisted before an object can be consumed by downstream tasks. To facilitate efficient fault recovery from the persisted lineage, we engage in upstream backup and parallel recovery as in Spark. In contrast to Spark's data parallel recovery, we conduct pipeline parallel recovery. We call this approach \textit{write-ahead lineage}.

The core idea of write-ahead lineage, shown in Algorithm 1, is simple: \textbf{as tasks process data partitions, the dynamically determined lineage is persisted to a transactional data store, the \textbf{GCS}\footnote{Global Control Store, inspired by Ray}, before the output can be consumed by downstream tasks}. When a task fails, this logged lineage can be consulted to recover from the failure by replaying data partitions and retrying tasks, similar to how Spark uses its statically determined lineage for recovery and how ACID database recover data from the write-ahead log \cite{zaharia2012resilient}. 

\begin{algorithm}
\caption{Write-ahead Lineage}
\begin{algorithmic}
\State $\text{Given task } \tau \text{ on worker } \omega \text{, with GCS } \mathcal{G}, \text{ where } \mathcal{G.L} \text{ stores }$
\State $\text{committed lineages, }\mathcal{G.T} \text{ stores outstanding tasks}$
\State $\mathcal{A} \gets \text{all data partitions pushed to } \omega $
\State $\mathcal{B} \gets \text{all possible inputs to } \tau $
\State $\mathcal{I} \gets \{ x \in \mathcal{A} \cap \mathcal{B} \mid x \in \mathcal{G.L}\}$
\If{$\mathcal{I} = \emptyset$}
    \State \textbf{return} \Comment{No inputs with committed lineage available}
\EndIf
\State $\text{Execute } \tau \text{, push results downstream}$
\State Store results locally on disk (upstream backup)
\If{push results failed}
    \State \textbf{return} \Comment{Downstream worker failure, do not commit}
\EndIf
\State $\text{Set } \tau \text{ to } \mathcal{I} \text{ in } \mathcal{G.L} \text{, remove } \tau \text{ from } \mathcal{G.T} \text{ in a single transaction.}$
\State \textbf{return} \Comment{Success}
\end{algorithmic}
\end{algorithm}

Intuitively, write-ahead lineage upholds the core invariant of lineage-based recovery: \textit{tasks consume only objects with committed lineage}. This ensures that a task's output stays the same after failure recovery, so channels that did not suffer failures do not have to be rewound. There are two classes of channels to consider here: those whose output is consumed by the failed task and those who consume the output of the failed task. For the first class, because all past outputs are backed up to local disk, outputs can be replayed.  Channels in the second class simply ignore the recovered task's re-transmitted output until the failed channel has recovered to the state before failure.\footnote{In systems based on Chandy-Lamport without lineage tracking, a failed worker can process inputs in different orders upon recovery, causing it to retransmit different messages than before failure. In Falkirk Wheel, this is called the ``no messages are duplicated" constraint. This typically results in expensive coordinated rollbacks of all channels to a globally consistent state in these systems, which we avoid \cite{gog2021falkirk, carbone2015apache}.}

There are two ways to enforce the core invariant: check lineage before consuming inputs or commit lineage before pushing outputs. Quokka adopts the former approach to minimize write transactions to the GCS, which lets the lineage be written as the last step of the algorithm. Quokka can then bundle this write with other writes to the GCS, like removing $\tau$ from the task queue and adding the next task in the channel, as a single transaction.

Only eventual consistency is required for the lineage. If a task does not immediately see a required input's lineage in the GCS, it will simply exit without being executed. The task will be tried again later and successfully execute when the lineage becomes visible.


\begin{figure}
    \centering
    \includegraphics[width=\linewidth,keepaspectratio]{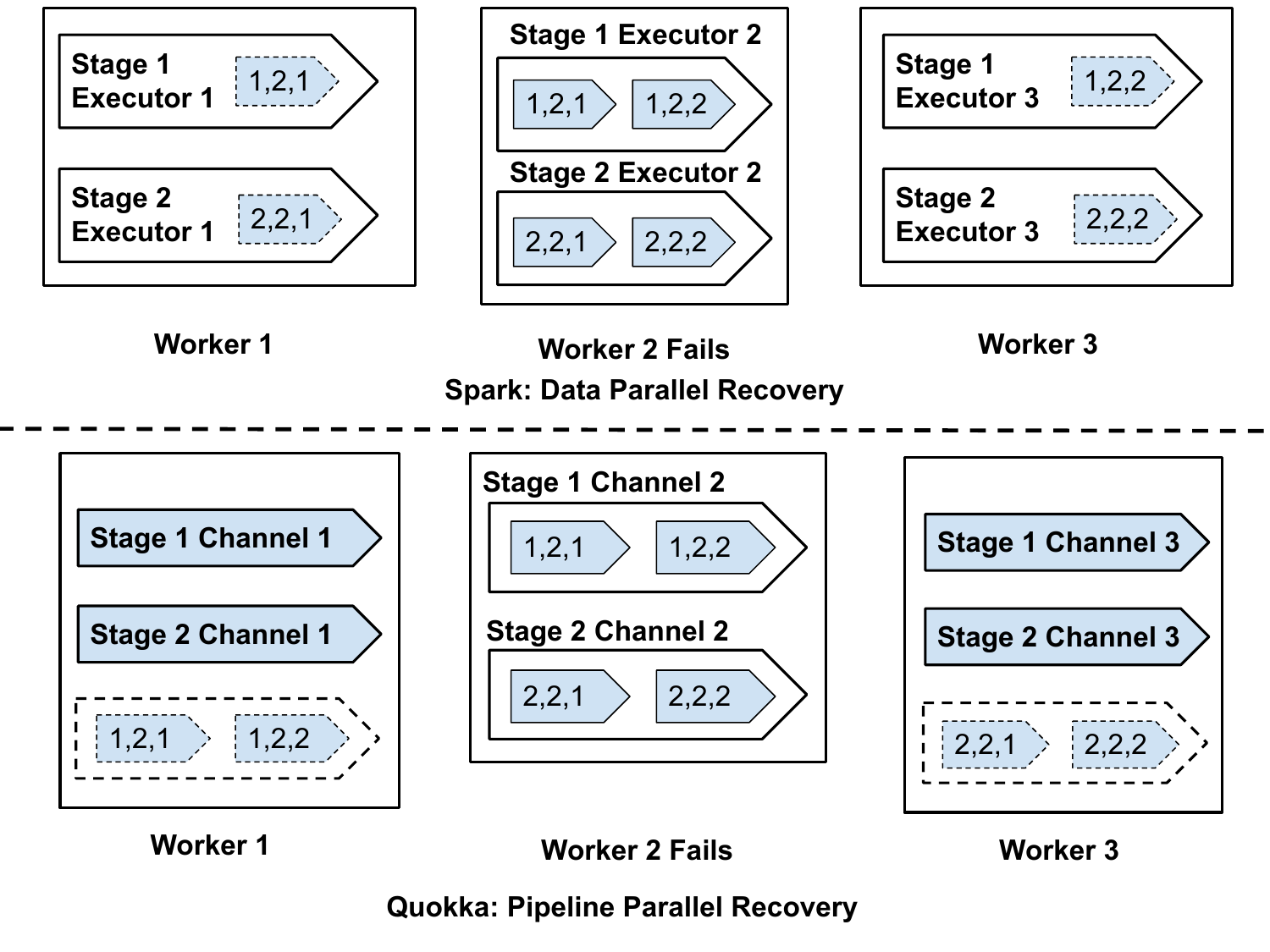}
    \caption{
        Spark employs data parallel recovery, where different tasks of the same stage are assigned to different workers. Quokka conducts pipelined parallel recovery, where different stages are assigned to different workers. We assume worker 2 fails and the dashed lines indicate recovery tasks.
        \label{fig:parallel-recover}
        \vspace{-0.5em}
    }
\end{figure}

\subsection{\textbf{Lineage Naming Scheme}} \label{sec:naming}

Recording lineage requires a naming scheme for tasks and their outputs. In Quokka, we introduce a naming scheme that allows a very succinct representation of the lineage to minimize logging overhead. The name of a task is a tuple of the form {\em (stage, channel, sequence number)}. The sequence number increases monotonically within each channel. A task's output has the same name as the task. Tasks must consume their inputs in order. As an example, in a query with two stages and two channels in each stage, a task in channel 1 in stage 2 could depend on outputs from tasks in either channel in stage 1. However, it must consume output from task (1,1,0) before task (1,1,1). In Quokka, we further restrict tasks to consume from one upstream channel at a time. A task decides at runtime how many task outputs from that channel to consume.

Under this execution model, a task's input requirement $\mathcal{B}$ can be described as a vector of length $C$, where $C$ is the number of upstream channels it could depend on. The $i^{th}$ element denote the number of consumed outputs from channel $i$ similar to a watermark. A task's lineage can be described with just two numbers, $i$ and $K$, which upstream channel it consumed from and how many outputs it consumed. This is much less information to log than a naive scheme where we assign unique names to all outputs in the system and keeping track of all input names for each task.


Our experiments indicate that write-ahead lineage's overhead in normal operation typically results from the disk writes needed for upstream backup. Quokka's upstream backup is similar to Spark, with the key difference being that in Quokka, multiple stages could be writing shuffle partitions at the same time, exerting higher disk pressure than Spark. We believe this design is reasonable as fast instance-attached NVMe SSDs are becoming more popular on public clouds, and network throughput used in shuffling data partitions will be saturated before disk throughput used in backing them up. The total amount of data stored for the entire job is the same between Quokka and Spark, as Spark typically maintains shuffle partitions of all stages.

\subsection{\textbf{Pipeline Parallel Recovery}} \label{sec:pipeline-parallel-recovery}

When a worker fails, the channels scheduled on it will lose their current active tasks, the associated state variables, and some cached data partitions. Quokka attempts to recover these channels to their previous state before failure by reconstructing lost partitions and state based on the logged lineage. 

Tasks in channels that do not contain state variables, typically input readers from object storage or stateless user defined functions, can be recovered in parallel across the cluster similar to Spark. However, within a failed channel with state variables, tasks must be reconstructed in sequence.

Even though tasks must be reconstructed in sequence \textit{within} a stateful channel, Quokka can still accomplish parallel recovery \textit{between} channels, as illustrated in Figure \ref{fig:parallel-recover}. In Quokka, if the query contains multiple stages (e.g. multi-way join), a failed worker contains many stateful channels that need to be reconstructed. These stateful channels belonging to different stages can be scheduled on different workers, in a pipelined parallel fashion. 

Compared to Spark where the degree of parallelism is proportional to the number of machines, the degree of parallelism here is proportional to the number of pipelined stages in the query. While Spark's scaling clearly wins at hundreds of machines, real production queries are commonly run on clusters of single digit sizes, on the same order of magnitude as the number of query stages \cite{vuppalapati2020building}.

\section{\textbf{Implementation}}\label{implementation}

We now describe how write-ahead lineage fits with other pieces of the query engine in Quokka. A simplified schematic of Quokka's architecture is shown in Figure \ref{fig:Quokka-arch}. 

\subsection{\textbf{Architecture}}

Quokka uses a cluster of worker machines, which might fail at any time, e.g. due to spot instance pre-emptions or Kubernetes pod evictions. Quokka also needs a head node, which could be one of the worker machines, or a separate instance. Like Spark, Quokka assumes the head node does not fail \cite{zaharia2016apache}, which can be achieved by using on-demand instances that cannot be preempted, or Kubernetes scheduling policies. In addition, we assume there is a client machine, which could be the user's laptop or another cloud instance, that submits jobs to this cluster. 

Quokka is implemented on top of Ray, a Python-based actor framework \cite{moritz2018ray}. Each physical worker is assigned \textit{TaskManagers} similar to Apache Flink \cite{carbone2015apache}. A TaskManager, implemented as a Ray actor, can be thought of as a thread pool that can be assigned tasks. The number of TaskManagers on a node configure inter-process parallelism, while the number of threads inside a TaskManager control threaded parallelism. Quokka also implements the coordinator as another Ray actor, which resides on the head node.

\begin{figure}
    \centering
    \includegraphics[width=\linewidth,keepaspectratio]{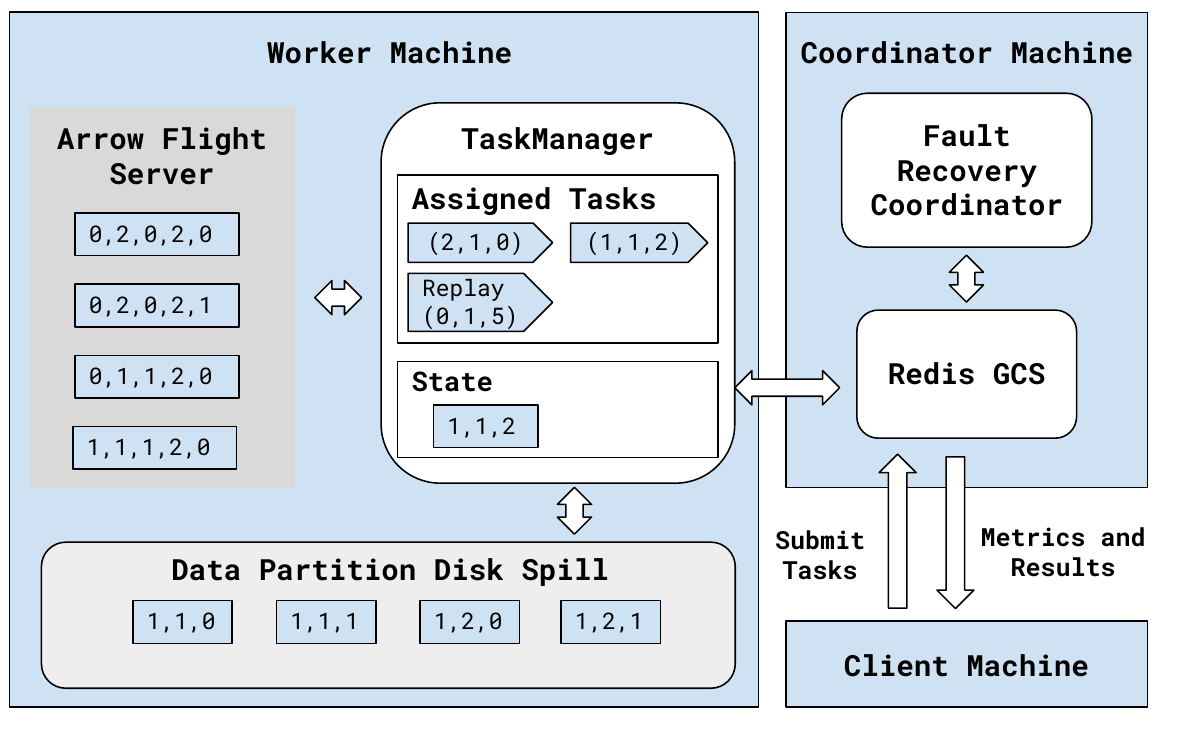}
    \caption{
        Quokka's architecture. Note that instead of having components communicate with each other through RPC calls, all coordination is done through the GCS. The client also communicates with the cluster through the GCS.
        \label{fig:Quokka-arch}
        \vspace{-0.5em}
    }
\end{figure}

A Quokka job consists of a sequence of stages. If the stage admits data parallelism, e.g. a join, then it can consist of multiple channels. A TaskManager is assigned one channel from each stage. Data parallel stages are thus parallelized across the entire cluster. For example, the TaskManager shown in Figure \ref{fig:Quokka-arch} is assigned channel 1 from stages 1 and 2. Tasks in a channel in stage 2 could depend on tasks in all channels in stage 1.

Task dependencies in Quokka are determined dynamically at runtime. While Quokka supports multiple scheduling strategies, in this paper we focus on a simple strategy: each task attempts to maximize the number of input batches it consumes. While this strategy may not be optimal, it is simple to understand and has strong performance. Quokka relies on DuckDB and Polars for single-node kernels such as join or filter, which generally perform better with larger batch size \cite{raasveldt2019duckdb, polars}. This strategy trades off maximizing single task efficiency with minimizing pipeline latency.

Quokka is a push-based query engine, where producer tasks push outputs to the workers hosting consumer tasks. We assume the mapping of channels to physical workers is static or maintained by a centralized lookup table. We believe write-ahead lineage can be applied to pull-based query engines as well, and do not make an argument for push or pull-based execution here.

All the TaskManagers on the same machine share access to an Apache Arrow Flight server, which manages zero-serialization data communication between different machines \cite{flight}. In Quokka a task pushes its outputs directly to the Arrow Flight servers of all its downstream consumer channels. We found the performance of this approach exceeds moving data via Ray's built-in object store and offers more flexibility in terms of handling disk spilling and chunking shuffle batches. 

All TaskManagers share access to the instance-attached disk for upstream backup of task outputs. Since the naming scheme described in Section \ref{sec:naming} ensures that the task outputs from different channels are named differently, there is no need for synchronization among different TaskManagers for writes.

\subsection{\textbf{Coordination through Transactions}}

The write-ahead lineage algorithm requires a persistent transactional data store, which we call the GCS. Quokka uses a Redis server on the head node to implement the GCS. Since we assume the head node does not fail during the job, anything logged to the Redis server is considered ``persisted". If head node failure is a concern, DynamoDB can be used instead.

The use of a GCS is inspired by the design of modern distributed systems like Kubernetes and Ray that offload the control plane to a data store such as etcd or Redis \cite{burns2016borg, moritz2018ray}. In addition to the lineage, the GCS holds the single source of truth for the execution state of the entire system in Quokka, such as the tasks assigned to each TaskManager and what data partitions are present on which machines. Individual TaskManagers are stateless and actively poll the GCS for tasks assigned to them to execute the write-ahead lineage algorithm in Algorithm 1. 

The coordinator periodically polls TaskManagers to see if any of them have failed. Once a failure is detected, the coordinator sets a control flag in the GCS. TaskManagers periodically poll this flag. If they see the flag is set, they abort their current tasks and wait. This barrier effectively implements a GCS-level lock to guarantee the coordinator exclusive read-write access to the GCS without potential conflicts. The coordinator then proceeds to schedule pipelined parallel recovery of tasks as described in Section \ref{sec:pipeline-parallel-recovery}.

The usage of a centralized GCS greatly simplifies the implementation of the coordinator. The coordinator's fault recovery routine simply updates the GCS with the tasks that need to be retried. In particular, it does not interact directly with TaskManagers. This separation has the key advantage that the coordinator does not have to assume the remaining TaskManagers are all alive, simplifying the handling of nested failures. We found this design to be much simpler and less error-prone than a traditional approach where different components of the system interact through RPC calls between the coordinator and TaskManagers, and allows us to avoid some SparkSQL fault recovery problems we encounter in practice on AWS EMR described in Section \ref{sec:fault-recovery}. 

A potential trade-off of a centralized GCS is performance and GCS memory footprint, especially when every task has to write to the GCS. Ray now uses the concept of Ownership to conduct distributed lineage tracking \cite{wang2021ownership}, where a task commits its lineage to other tasks who have a claim to its results. However, with the optimized task naming scheme described in Section \ref{sec:naming}, both the GCS logging overhead and its memory footprint become negligible in Quokka.

\subsection{\textbf{Failure Recovery}}\label{ft-implementation}

Quokka's approach to failure recovery is motivated by Kubernetes' philosophy of \textit{reconciliation}. When a failure occurs, the GCS now contains inconsistent information, such as tasks assigned to failed workers. During fault recovery, the coordinator updates it to a consistent state satisfying the following constraints: 
\begin{itemize}
  \setlength\itemsep{0em}
    \item lost tasks are rescheduled on live TaskManagers.
    \item all the input data partitions needed for any existing or rescheduled task will be replayed or recomputed.
\end{itemize}

The algorithm used by the coordinator in Quokka is shown in Algorithm 2 with simplifications\footnote{In Algorithm 2, we assume there are no lost replay or input tasks, though Quokka handles those as well.}. It implements pipeline parallel recovery described in Section \ref{sec:pipeline-parallel-recovery} by assigning different rewound stateful channels to different workers.

\begin{algorithm}
\caption{Failure Recovery Algorithm}
\begin{algorithmic}
\State Assume GCS $\mathcal{G}$, where $\mathcal{G.L}$ stores committed lineages, $\mathcal{G.T}$ stores outstanding tasks
\State $\mathcal{A} \gets \text{the set of all tasks assigned to the failed worker}$
\State $\mathcal{R} \gets \{(\tau.\text{stage}, \tau.\text{channel}) \mid \tau \in \mathcal{A}\} \text{ (Rewind requests)}$
\For{\textbf{each} $(stage, channel)$ \textbf{in reverse topological order}}
    \If{$(stage, channel) \in \mathcal{R}$}
        \State Identify required inputs $\mathcal{I}$ for $(stage, channel)$ from lineages of $channel$ outputs in $\mathcal{G.L}$
        \For{\textbf{each} data partition $(stage, channel, seq)$ \textbf{in} $\mathcal{I}$}
            \State If exists, add replay task to the owner worker
            \State Else if $stage$ is input, add input task to any node
            \State Else, add $(stage, channel, 0)$ to $\mathcal{R}$
        \EndFor
    \EndIf
\EndFor
\For{\textbf{each} $(stage, channel, 0)$ \textbf{in} $\mathcal{R}$}
    \State Remove $(stage, channel, seq)$ from $\mathcal{G.T}$
    \State Assign $(stage, channel, 0)$ to a random worker in $\mathcal{G.T}$
\EndFor
\end{algorithmic}
\end{algorithm}

\begin{figure}
    \centering
    \includegraphics[width=\linewidth,keepaspectratio]{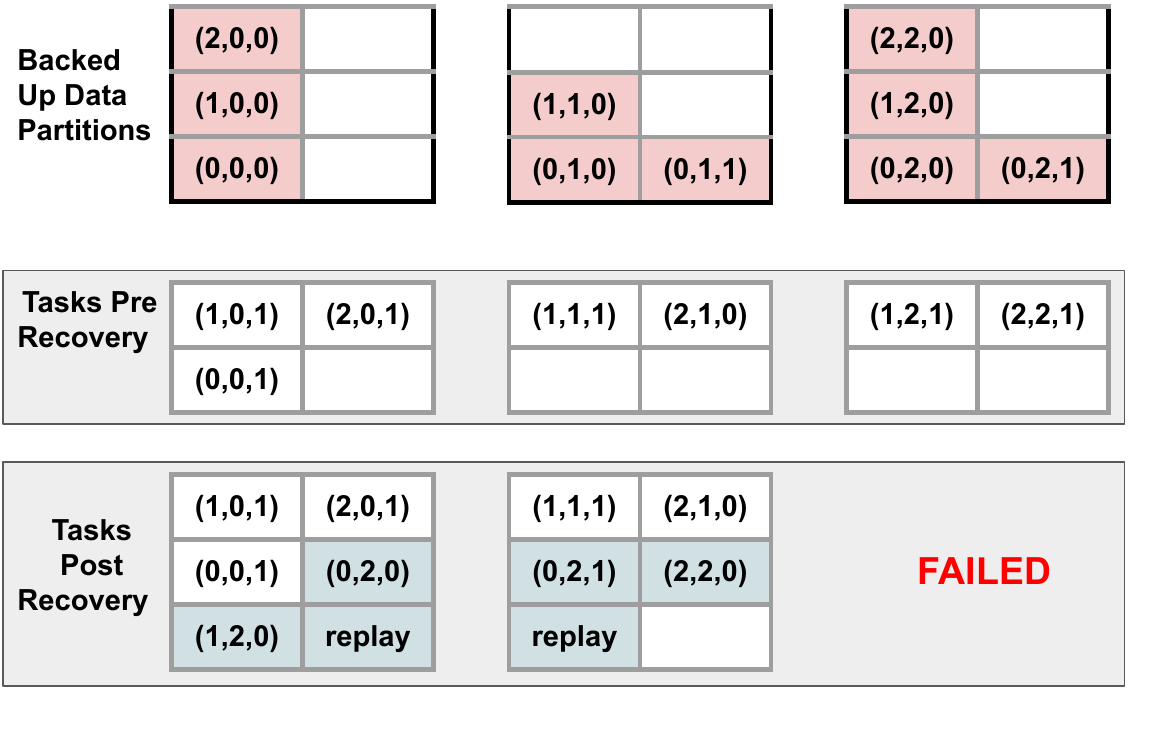}
    \caption{
        An example fault recovery procedure when one out of three workers fail. Pink shade represents data partitions that have been generated by past tasks and stored on the TaskManager. Recovery tasks are shaded in light blue.
        \label{fig:failure}
        \vspace{-0.5em}
    }
\end{figure}

A concrete example is shown in Figure \ref{fig:failure} for a Quokka application with three stages. Stage 0 is stateless and stage 1 and stage 2 make use of state variables. After a TaskManager fails, the coordinator will first reschedule all its tasks,  (1,2,1) and (2,2,1) in this example, to other live TaskManagers.  These tasks are restarted from the initial state, so the tasks that need to be launched are (1,2,0) and (2,2,0), which can be relaunched on different machines.

The coordinator now traverses the stages in reverse topological order and checks if the required data partitions for the relaunched tasks are still present. If so, e.g. (0,0,0), (0,1,0) ... , replay tasks are pushed to TaskManagers that hold them. If not, the data partition must be regenerated by rewinding other channels. This is typically due to tasks on the failed machine depending on data partitions held by the same machine, i.e. (0,2,0) and (0,2,1). 

Importantly, when a rewound task such as (1,2,0) or (2,2,0) is ``retracing its footsteps'', it is no longer free to dynamically choose its input data partitions. Instead, the GCS is consulted to supply it with the exact lineage to regenerate each output data partition, which ensures the rewound channel regenerates the same outputs as before the failure. 

As mentioned in Section \ref{ft-theory}, Quokka can engage in pipelined parallel recovery between channels, so (1,2,0) and (2,2,0) can be rescheduled to be recovered on different workers. (0,2,0) and (0,2,1) can be rescheduled on different workers even though they belong to the same channel, since there's no state dependency between them. In total four data partitions need to be reconstructed in this failure scenario, which corresponds to the total number of data partitions stored on the failed machine. 




\begin{figure*}
    \centering
    \includegraphics[width=\linewidth]{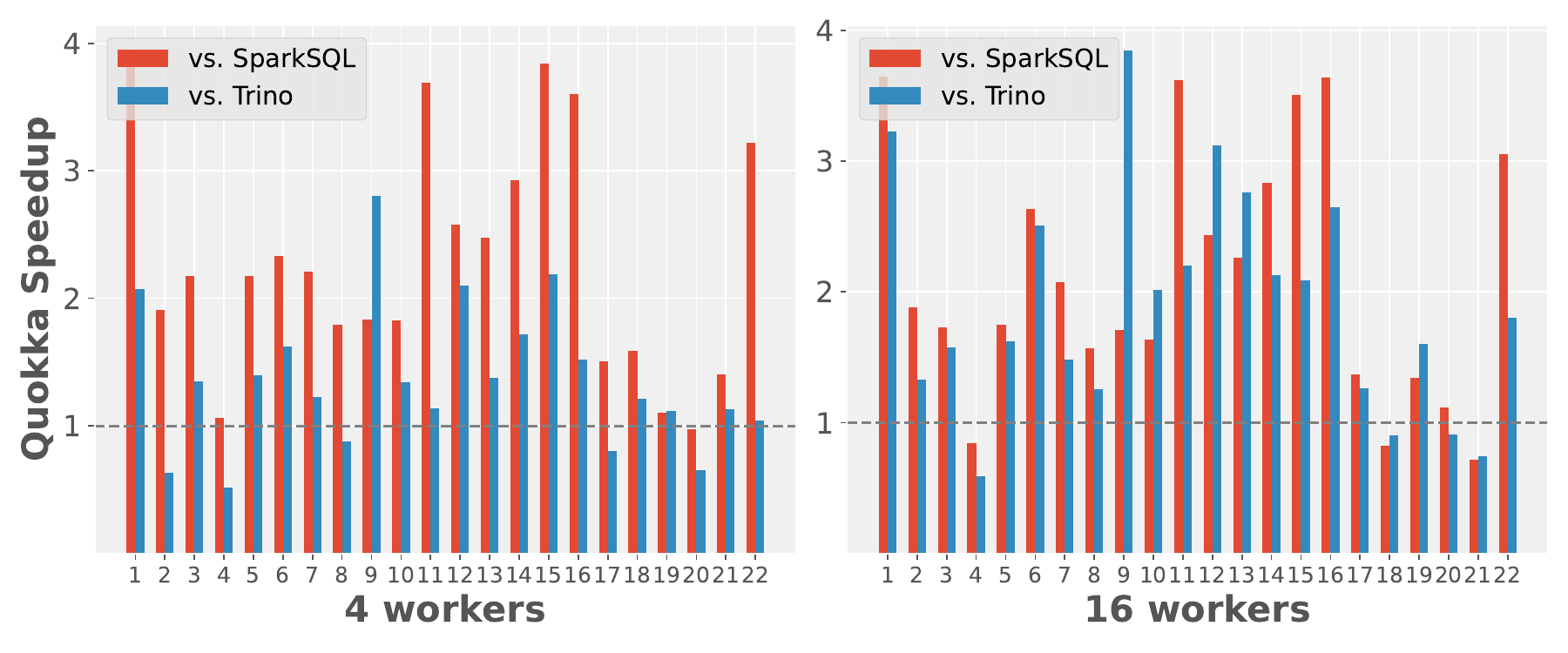}
    \caption{
        Comparing the performance of fault-tolerant data processing systems (Trino with FT, SparkSQL and Quokka) on the TPC-H queries on a) 4-worker cluster and b) 16-worker cluster. Quokka outperforms Trino and SparkSQL in most cases.
        \label{fig:systems}
        \vspace{-1em}
    }
\end{figure*}

\section{\textbf{Evaluation}}

We test Quokka's performance, fault tolerance and scalability on the full TPC-H benchmark (scale factor 100) with input in Parquet format stored on AWS S3. We then select 8 representative queries in three different categories: 

\begin{itemize}
    \item \textbf{I}: simple aggregations (1, 6)
    \item \textbf{II}: simple pipelined joins (3, 10) 
    \item \textbf{III}: queries with multiple join pipelines (5, 7, 8, 9)
\end{itemize}

We perform detailed ablation studies of different design choices and measure fault recovery performance on these representative queries. These queries are chosen because they contain mostly just one join tree, reducing confounding variables on performance.

Quokka is run on a Ray cluster with Ray version 2.4 on AWS EC2 on-demand instances. We use two cluster configurations. The first configuration uses four r6id.2xlarge worker machines. The second uses 16 r6id.xlarge machines. An r6id.2xlarge instance has 8 vCPUs, 64GB of RAM ad 474GB of instance attached NVMe SSD. An r6id.xlarge instance has exactly half of those resources.

For comparison, we benchmark SparkSQL 3.3 and Trino 398 on AWS EMR 6.9.0 on the same cluster configurations. AWS EMR configures Spark to also use NVMe SSDs for potential spilling. We further optimize the network and shuffle retry configurations of SparkSQL to start fault recovery in two seconds, instead of the default two minutes on AWS EMR, to match the behavior of Quokka. Trino is benchmarked with and without fault tolerance by HDFS spooling. SparkSQL is fault tolerant by default. Before running the queries, \texttt{ANALYZE} commands are run for both SparkSQL and Trino to ensure cardinality-based optimizations are enabled.

Unless otherwise noted, all timing results are from the mean of three independent measurements, with standard deviation shown as error bars when applicable.

\subsection{\textbf{Quokka vs. Trino vs. SparkSQL}} In Figure \ref{fig:systems}, we compare Quokka's performance to Trino with spooling-based fault tolerance and SparkSQL on the full TPC-H benchmark. We see that for most queries across both cluster configurations, Quokka is the most performant among all three query engines. Compared to Trino, Quokka achieves 25\% geometric mean speedup on the 4-worker cluster and 70\% on the 16-worker cluster. Compared to SparkSQL, Quokka achieves 2.1x geometric mean speedup on the 4-worker cluster and 1.9x on the 16-worker cluster. 

It is important to note that a lot of factors could contribute to these results. All three systems employ different kernels to implement SQL operators such as join and filter, libraries for networked communication and task scheduling systems. However, these results do indicate that Quokka's implementation is competitive with state-of-the-art data processing systems. We attribute Quokka's speedup over Spark mostly to blocking vs pipelined execution and Quokka's speedup over Trino to Trino's high spooling overhead.

We note that Quokka's performance against SparkSQL and Trino is worst for complicated queries that contain nested subqueries and might require materialization of intermediate results (e.g. 2, 4, 20, 21) and better for simpler queries like 8, 9 or 12 that just contain one join tree. The reason for this discrepancy is Quokka currently must perform expensive global synchronization between pipelines, making complicated queries that contain multiple pipelines slow. In addition, Quokka's implementation of semi-joins and anti-joins, required to unnest subqueries, are not yet very efficient.

Quokka's advantage against SparkSQL and Trino is maintained on the 16-worker cluster compared to the 4-worker cluster, suggesting that Quokka's design is scalable to larger cluster sizes.

\begin{figure*}
    \centering
    \includegraphics[width=\linewidth,keepaspectratio]{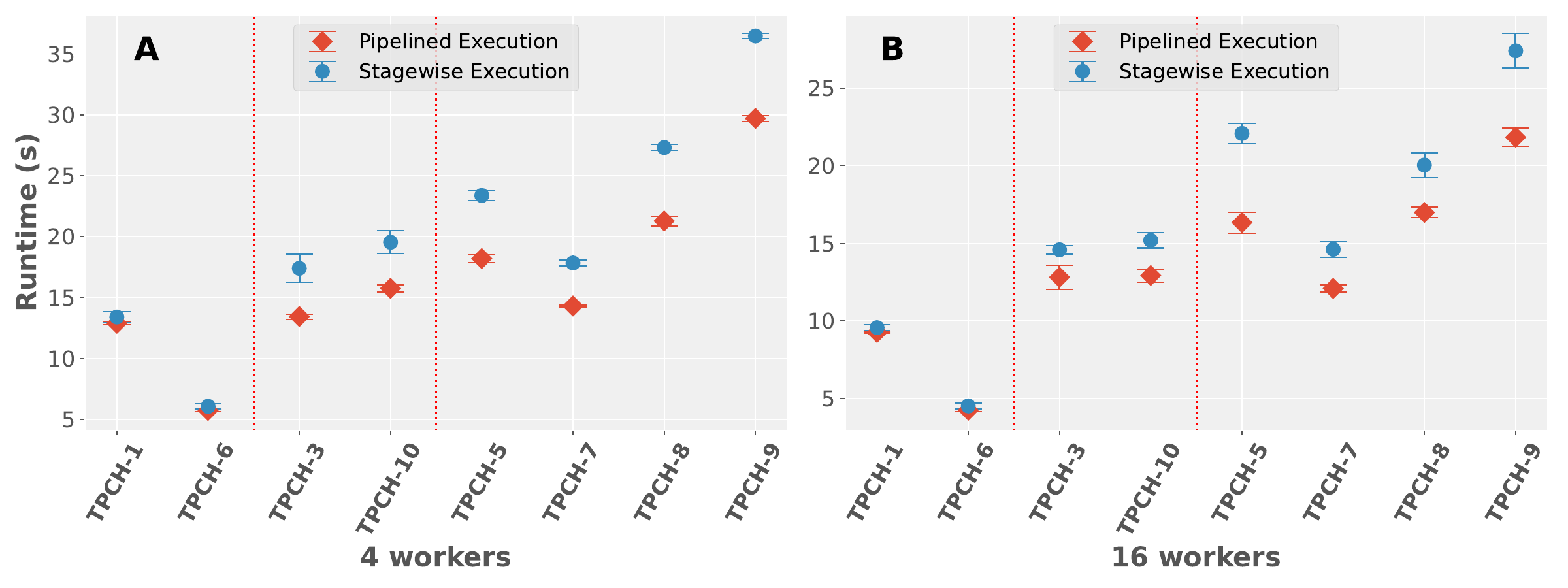}
    \caption{
        Pipelined Quokka vs Stagewise (blocking) Quokka execution times on the TPC-H queries on the a) 4-worker cluster and b) 16-worker cluster. Pipelined execution outperforms in all cases.
        \label{fig:blockvsdynamic}
        \vspace{-0.5em}
    }
\end{figure*}

\begin{figure*}
    \centering
    \includegraphics[width=\linewidth,keepaspectratio]{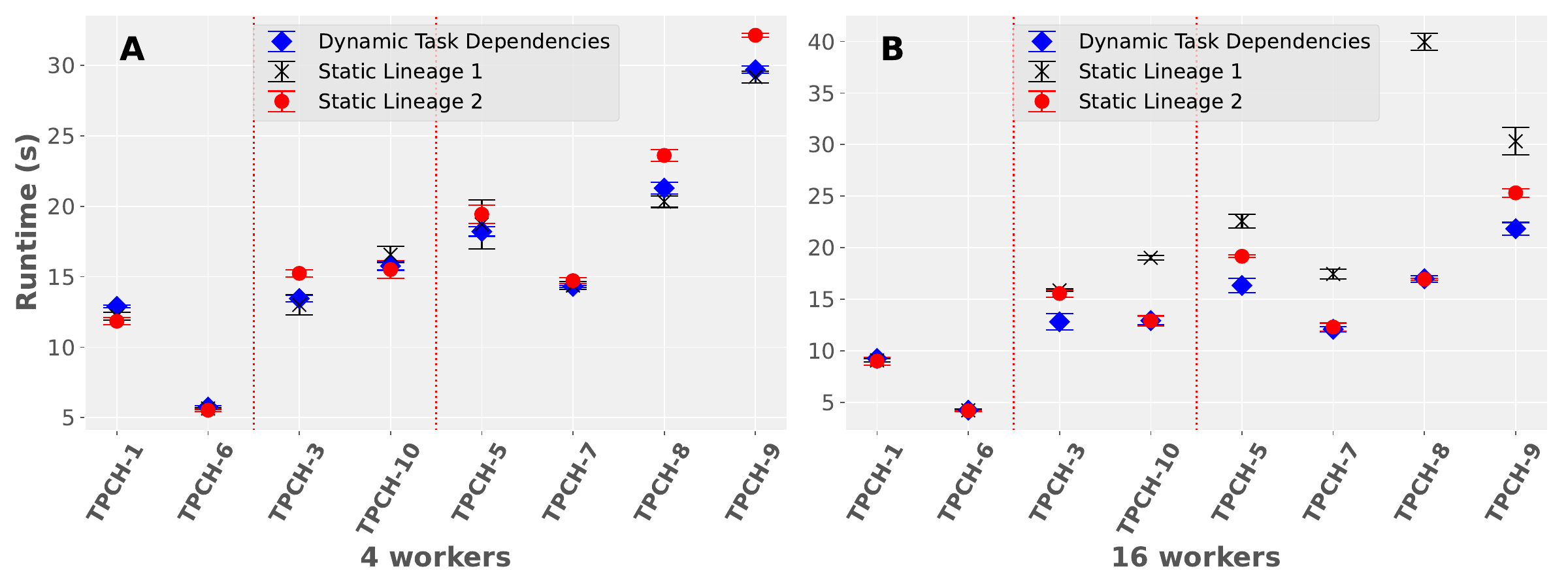}
    \caption{
        Performance of Quokka with dynamic task dependencies vs. two different static lineage strategies on the a) 4-worker cluster and b) 16-worker cluster. Strategy 1 (batch size 8) outperforms strategy 2 (batch size 128) on the 4-worker cluster but greatly underperforms on the 16-worker cluster. Enabling dynamic task dependencies allows Quokka to match the better performing static strategy in most cases.
        \label{fig:dynamic-vs-static}
        \vspace{-0.5em}
    }
\end{figure*}

\subsection{\textbf{Why Dynamic Pipelined Execution?}}

We now show that dynamic pipelined query execution leads to significant performance gains compared to both stagewise execution and pipelined query execution with static task dependencies to motivate the need for a fault tolerance algorithm specifically designed for dynamic pipelined query engines. 

\subsubsection{\textbf{Pipelined vs Blocking Execution}} We modify Quokka to execute in a stage-wise fashion similar to SparkSQL and examine its performance degradation. Results for both cluster sizes are shown in Figure \ref{fig:blockvsdynamic}. 

Across the eight selected queries across the two different cluster setups pipelined execution consistently outperforms stagewise execution. The speedups are especially significant for queries in category III, which involve multiple joins that can be pipelined. On queries 1 and 6 in category I where the query consists of basically just the read stage, the pipelined execution does not improve runtime, as expected.

Overall, pipelined execution leads to 26\% geometric mean speedup on the queries in categories II and III on the 4-worker cluster and 22\% speedup on the 16-worker cluster. On queries with deep join trees like query 8, the speedup could be as large as 28\%. 

\subsubsection{\textbf{Dynamic vs Static Lineage}} If we could achieve good performance with static task dependencies determined before query execution, we do not have to bother with logging the lineage during query execution. 

\begin{figure*}
    \centering
    \includegraphics[width=\linewidth,keepaspectratio]{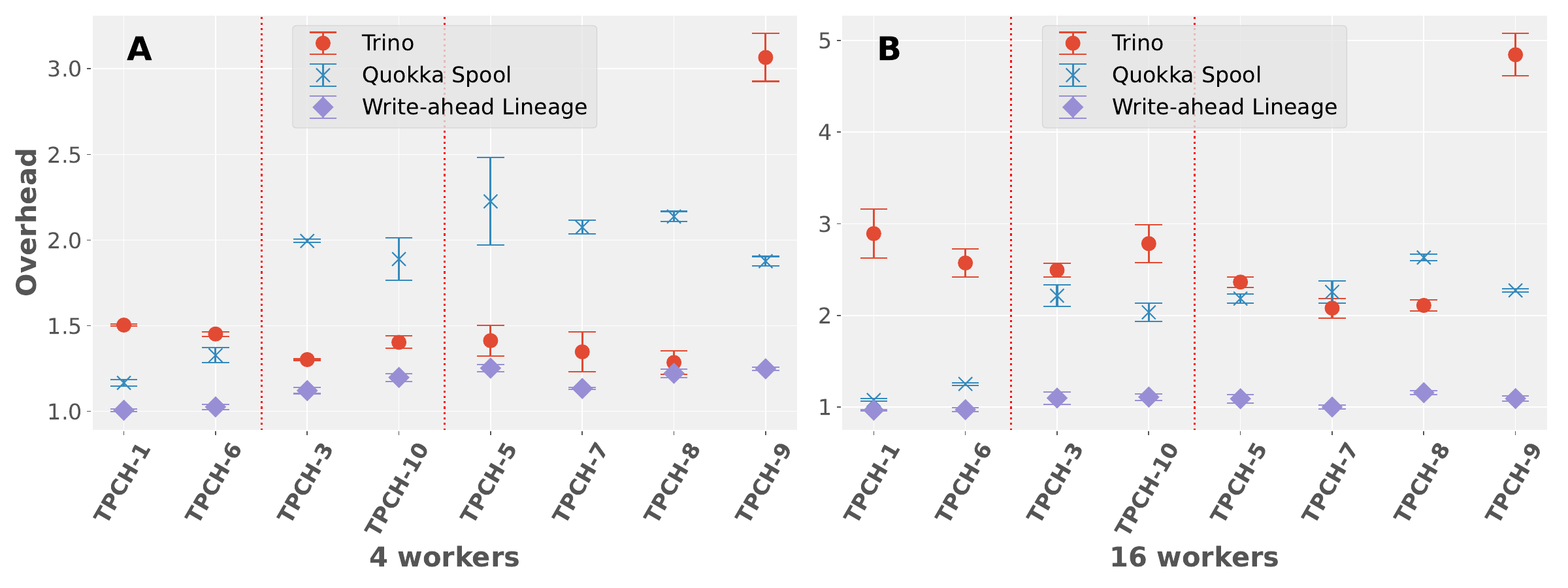}
    \caption{
        Trino's HDFS spooling fault tolerance overhead, Quokka S3 spooling overhead and write-ahead lineage overhead on the a) 4-worker cluster and b) 16-worker cluster. Overhead of 1 means no overhead.
        \label{fig:spool-overhead}
        \vspace{-0.5em}
    }
\end{figure*}

In a \textit{static lineage} strategy, a task consumes a fixed number of input data partitions at a time. If this number is too small, a high volume of smaller partitions will be transmitted across the network, diminishing network efficiency. However, if this number is too big, effective pipelining cannot occur, and the system effectively executes in a stage-wise fashion similar to SparkSQL. It is very difficult to statically choose this number correctly in practice, since the sizes of data partitions can depend on the size of the cluster, data distributions and join and filter selectivity. 

To demonstrate this difficulty, we show the performance of two static lineage strategies across the selected queries on the 4-worker and 16-worker clusters in Figure \ref{fig:dynamic-vs-static}. In the first strategy, stateful operators batch together 8 input data partitions for each execution. In the second strategy, they batch together 128.

Similar to the previous experiment, performance differences between these strategies are not apparent for the simple queries in category I. However, differences become more significant with join queries in category II and more so with more complex joins in category III. We see that on the 4-worker cluster, a batch size of 8 is clearly superior to a batch size of 128, while the reverse is true on the 16-worker cluster. 

On the 16-worker cluster, a batch size of 8 causes very small partitions to flow through the system, causing a marked decrease in CPU utilization and network I/O efficiency during shuffles. On the 4-worker cluster, the partition slices shuffled were larger due to the reduced parallelism, causing batch size 8 to instead outperform batch size 128. 

Allowing tasks to have dynamic task dependencies allow Quokka to achieve similar or better performance than the better of the two static lineage strategies in both cluster settings for most queries. 



\subsection{\textbf{Write-ahead Lineage Overhead}} In this section, we benchmark the overhead imposed by Quokka's write-ahead lineage algorithm during normal execution and compare it to spooling based options. We turn Trino's fault tolerance off to measure the overhead added by its HDFS spooling in Figure \ref{fig:spool-overhead}.\footnote{The overhead is defined by the ratio of ratio of runtimes with and without fault tolerance. A value of 1 means there is no overhead.}

Across the selected queries, Trino's spooling adds a geometric mean 1.5x overhead on the 4-worker cluster and 2.7x overhead on the 16-worker cluster, reaching up to 4.8x in the case of query 9. The overhead is considerably worse on the 16-machine cluster compared to the four-machine cluster. We believe this is because as the data partitions that need to be spooled to HDFS become smaller, HDFS efficiency markedly decreases. We also experimented with S3-based spooling for Trino, which led to much worse results. 

We also implemented S3-based spooling in Quokka and observed similar overhead to Trino, as shown in Figure \ref{fig:spool-overhead}. Note Quokka's spooling overhead is minimal for the two queries in category I since Quokka's aggregation pushdown makes the spooled data size insignificant. It appears Trino does not perform this optimization. Quokka's spooling overheads are similar for simple joins in category II and complex joins in category III since most of the spooled data comes from the lineitems table, referenced by all the queries. 

In comparison, the overhead of Quokka's write-ahead lineage strategy is an order of magnitude better than the spooling options, only 15\% on the 4-worker cluster and 6\% on the 16-worker cluster. Like Spark, Quokka backs up partitions unreliably in the worker's local disk instead of RAM to save memory. However these local disk writes are a lot more efficient than networked HDFS or S3 writes, and can typically be hidden by computation and network IO.

\begin{figure*}
    \centering
    \includegraphics[width=\linewidth,keepaspectratio]{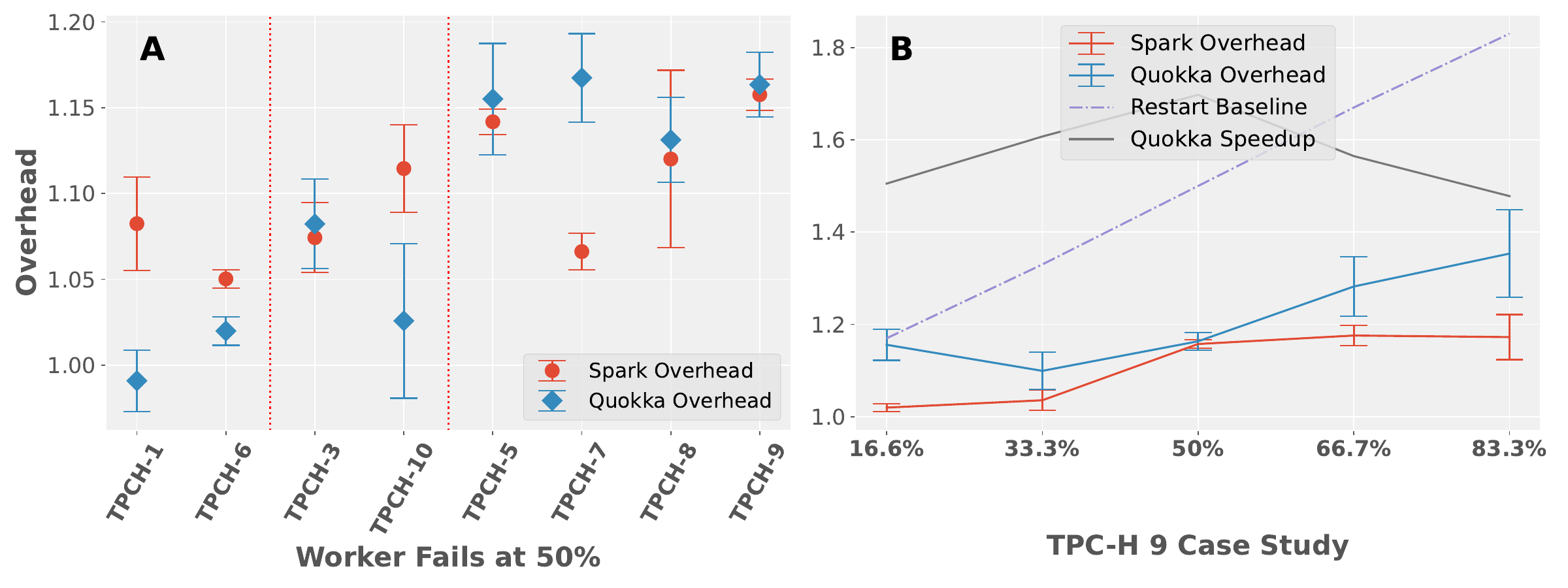}
    \caption{
        Quokka vs SparkSQL's fault recovery behavior. a) Quokka vs Spark fault recovery performance on the 16-worker cluster where a random worker is killed at 50\% query completion during each query. b) A case study for TPC-H 9 where worker dies at varying points during the execution. We also show Quokka's end-to-end speedup over Spark on the same y-axis scale.
        \label{fig:ft}
        \vspace{-1em}
    }
\end{figure*}

Compared to Spark, Quokka also needs to consistently log the lineage of each spilled partition, which currently happens via the Redis GCS on the head node. We find this cost to be negligible in our benchmark as optimizations described in Section \ref{sec:naming} greatly simplified the lineage. Virtually all the overhead results from the disk writes.

In addition to spooling, we also benchmarked Quokka with custom checkpointing strategies to S3. Even with incremental checkpointing, we observe severe overhead in normal operation. The biggest overheads come from operators whose state increases over time, like building the hash table for a shuffle hash join. While the exact overhead depends on the checkpointing interval, at any reasonable interval that's useful for recovery, it performs much worse than spooling to S3.


\subsection{\textbf{Fault Recovery Performance}} \label{sec:fault-recovery}

We now compare Quokka's fault recovery performance compared to SparkSQL. Instead of SparkSQL's data parallel recovery, Quokka engages in pipelined parallel recovery as described in Section \ref{sec:pipeline-parallel-recovery}.

The first fault recovery experiment consists of running each of our representative queries on the 16-worker cluster. A worker machine is killed halfway through the query based on its normal execution runtime. The fault recovery overhead, defined by total runtime with failure divided by normal runtime without failure, is shown in Figure \ref{fig:ft}a for both systems. 

We observe that Quokka and SparkSQL have similar recovery overhead, with Quokka's overhead better by a geometric mean of 1\%. We see that for both SparkSQL and Quokka, simpler queries in Category I tend to have lower recovery overhead compared to more complicated joins in Categories II and III. While Quokka is faster than SparkSQL at fault recovery in Category I, it is slightly slower than SparkSQL in Category III. Note that in every case, we significantly outperform the baseline of just restarting the query from scratch on the remaining workers, which corresponds to an overhead of 1.5x.

In Figure \ref{fig:ft}b, we show a case study on TPC-H query 9 where we show SparkSQL and Quokka's fault recovery performance when the query experiences a failure at different points throughout the query. As expected, Quokka incurs higher fault recovery overhead if the failure occurs late in the query, as there is more work to be redone. SparkSQL exhibits the same behavior. However in all cases, Quokka and SparkSQL's fault recovery performances significantly beat the simple baseline of restarting the query from scratch after failure. Even though Quokka has more recovery overhead, it still outperforms Spark end-to-end with the failure in all cases since it is much faster in normal execution.

We notice that despite SparkSQL's usual good fault recovery performance, it occasionally fails to recover by continuing to make RPC requests to the dead worker, which is a known problem for open-source Spark on AWS EMR \cite{spark-fail}. Our results only included trials where this problem is not encountered, which strictly improved SparkSQL's fault recovery results. Quokka's choice to communicate only through the GCS to avoid all direct RPCs between different components preclude it from this class of problems.

\begin{figure*}
    \centering
    \includegraphics[width=\linewidth,keepaspectratio]{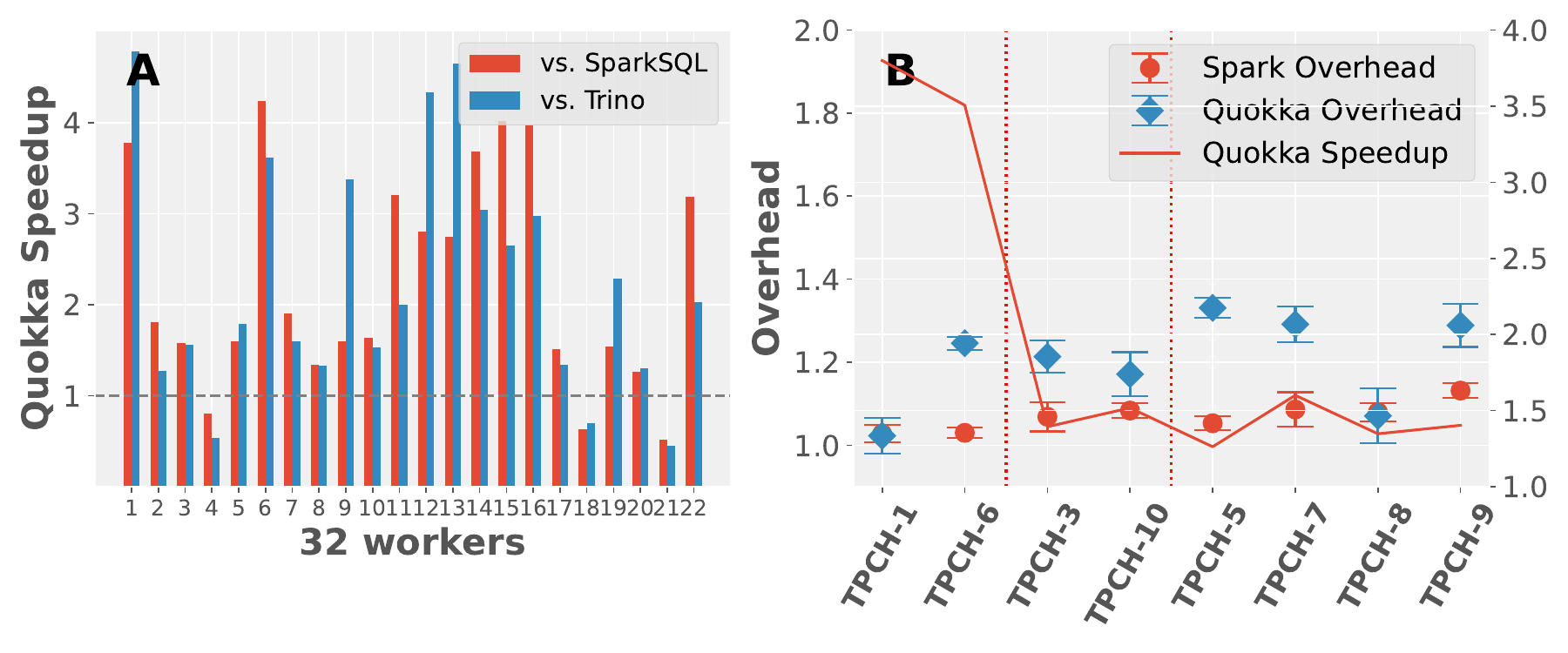}
    \caption{
        Experiments on 32 workers in terms of a) normal execution performance without failures compared to Spark and Trino with FT and b) fault recovery overheads where a random worker is killed at 50\% completion. Quokka is still faster end-to-end on each query compared to SparkSQL (right y-axis).
        \label{fig:32}
        \vspace{-1em}
    }
\end{figure*}

\subsection{\textbf{Scalability}}

\subsubsection{\textbf{Performance}}
We test the scalability of Quokka and write-ahead lineage with 32 rd6id.xlarge workers on the TPC-H benchmark queries with the same dataset. Figure \ref{fig:32}a shows the speedup Quokka achieves vs. Spark and Trino in this setting. The speedup profile across the queries is largely similar to the 4-worker and 16-worker settings. Quokka achieves geomean 1.92x speedup over Spark and 1.86x over Trino. Quokka's speedup over SparkSQL is stable while its speedup over Trino improves with the number of machines, confirming our observations in Section VIC that Trino's spooling overhead gets worse with the number of machines.

\subsubsection{\textbf{Fault Recovery}} \label{sec:fault-recover-scale}
Figure \ref{fig:32}b repeats the experiment shown in Figure \ref{fig:ft}a, where a worker machine is killed 50\% of the way through a representative query, for the 32-worker setting. We see that compared to the 16-worker setting, the recovery performance of Quokka deteriorates compared to Spark. On average, Quokka has 12\% worse geomean recovery overhead in this case, vs. 1\% better in the 16-worker setting. However, even though Quokka has more recovery overhead, it both outperforms the restart baseline (1.5x overhead) in all cases and outperforms Spark end-to-end with the failure in all cases due to its faster normal execution performance.

Quokka's degraded fault recovery performance at 32 nodes can be understood from the discussion in Section \ref{sec:pipeline-parallel-recovery}. Unlike Spark's data parallel recovery, pipelined parallel recovery only leverages parallelism up to the number of stages in the query, not the number of workers in the cluster. As a result, increasing the worker count from 16 to 32 improves Spark's recovery performance but not Quokka's. Our design point exploits the fact that  it is rare to see more than 16 nodes in practice, even for very large workloads \cite{vuppalapati2020building}.




\section{\textbf{Discussion and Conclusion}}

In this paper, we present write-ahead lineage, a novel fault tolerance mechanism for pipelined query engines with low overhead and fast recovery times. We showcase its implementation in a real query engine, Quokka, achieving 1.9x speedup against SparkSQL and 1.7x against Trino in normal operation on a 16-machine cluster on TPC-H, while matching SparkSQL in fault recovery performance. On 32 machines, Quokka maintains its strong performance while degrading slightly in fault recovery performance due to the reasons explained in Section \ref{sec:fault-recover-scale}. Ablation studies shown in Figure 7, 8 and 9 confirm Quokka's dynamic pipelined execution brings concrete performance benefits while write-ahead lineage incurs an order of magnitude less overhead compared to spooling-based alternatives.

\subsection{\textbf{Data vs Compute Fault Tolerance}}



Pipelined query engines were originally designed to be part of a database server. When a machine failed, the first concern was preventing data loss and minimizing system downtime, not recovering  a transient user query \cite{lamb2012vertica}. Intra-query fault tolerance became relevant in the era of decoupled compute and storage. Cloud storage now typically guarantees many nines of data durability, resolving concerns of data loss. On the other hand, computation is increasingly conducted by ephemeral resources that might fail or be pre-empted at any time.

In this setting, it makes sense to study how to recover from computation failures while assuming the input data is persistent and replayable, leading to popular fault-tolerant systems such as MapReduce and Spark \cite{dean2008mapreduce, zaharia2016apache}. However, their stagewise design leads to inefficiencies in query execution. Pipelined query engines such as Trino have attempted to add fault tolerance, though its spooling based approach has high overhead \cite{trino-ft}. Hosted query engines such as Snowflake are also fault tolerant to the user, but they simply restart failed queries under the hood \cite{snowflake-ft}.

In this paper, we adapt age-old write-ahead logging techniques used for data recovery to tackle the new problem of computation failure recovery for pipelined query engines.



\subsection{\textbf{Design Motivations and Novelty}}

Fault tolerance is a well-studied field with multiple established techniques such as lineage, spooling and checkpointing. While pipelined systems like Trino or StreamScope typically adopt a combination of spooling and checkpointing to achieve fault tolerance, we find these techniques cause high overheads in normal execution, as shown in Figure \ref{fig:spool-overhead} \cite{trino-ft, lin2016streamscope, kreps2011kafka}.

Quokka's write-ahead lineage combines persistent lineage logging with upstream backup, similar to Spark. However, important differences exist due to Quokka's novel setting of pipelined query execution with dynamic task dependencies. In contrast to Spark's static lineage, write-ahead lineage logs the lineage as it is determined during query execution. We show this imposes negligible overhead in normal execution in Figure \ref{fig:spool-overhead}. Instead of Spark's data parallel recovery, Quokka adopts pipeline parallel recovery, whose degree of parallelism scales with the number of stages in the pipeline instead of the number of workers. We show that Quokka has comparable fault recovery overhead to Spark on the most common workloads with up to 16 workers, only tailing off at 32 nodes while still maintaining an end-to-end performance improvement because of the benefits of pipelined execution.




\subsection{\textbf{Implementation}}



We have open sourced Quokka on Github\footnote{https://github.com/marsupialtail/quokka} to facilitate data systems research and applications. It supports a DataFrame API similar to Spark and Polars, and has already been used to support emerging data engineering applications like vector search on data lakes\cite{dataaisummit_vectordatalakes}. Even though this paper focuses on Quokka, we believe write-ahead lineage can easily be added to any distributed pipelined query engine, where a distributed key-value store like DynamoDB or FoundationDB can be used to track metadata \cite{dageville2016snowflake, zhou2021foundationdb, sivasubramanian2012amazon}.

\bibliographystyle{IEEEtran}
\bibliography{main.bib}

\begin{thebibliography}{10}
\providecommand{\url}[1]{#1}
\csname url@samestyle\endcsname
\providecommand{\newblock}{\relax}
\providecommand{\bibinfo}[2]{#2}
\providecommand{\BIBentrySTDinterwordspacing}{\spaceskip=0pt\relax}
\providecommand{\BIBentryALTinterwordstretchfactor}{4}
\providecommand{\BIBentryALTinterwordspacing}{\spaceskip=\fontdimen2\font plus
\BIBentryALTinterwordstretchfactor\fontdimen3\font minus \fontdimen4\font\relax}
\providecommand{\BIBforeignlanguage}[2]{{%
\expandafter\ifx\csname l@#1\endcsname\relax
\typeout{** WARNING: IEEEtran.bst: No hyphenation pattern has been}%
\typeout{** loaded for the language `#1'. Using the pattern for}%
\typeout{** the default language instead.}%
\else
\language=\csname l@#1\endcsname
\fi
#2}}
\providecommand{\BIBdecl}{\relax}
\BIBdecl

\bibitem{armbrust2015spark}
M.~Armbrust, R.~S. Xin, C.~Lian, Y.~Huai, D.~Liu, J.~K. Bradley, X.~Meng, T.~Kaftan, M.~J. Franklin, A.~Ghodsi \emph{et~al.}, ``Spark sql: Relational data processing in spark,'' in \emph{Proceedings of the 2015 ACM SIGMOD international conference on management of data}, 2015, pp. 1383--1394.

\bibitem{melnik2010dremel}
S.~Melnik, A.~Gubarev, J.~J. Long, G.~Romer, S.~Shivakumar, M.~Tolton, and T.~Vassilakis, ``Dremel: interactive analysis of web-scale datasets,'' \emph{Proceedings of the VLDB Endowment}, vol.~3, no. 1-2, pp. 330--339, 2010.

\bibitem{sethi2019presto}
R.~Sethi, M.~Traverso, D.~Sundstrom, D.~Phillips, W.~Xie, Y.~Sun, N.~Yegitbasi, H.~Jin, E.~Hwang, N.~Shingte \emph{et~al.}, ``Presto: Sql on everything,'' in \emph{2019 IEEE 35th International Conference on Data Engineering (ICDE)}.\hskip 1em plus 0.5em minus 0.4em\relax IEEE, 2019, pp. 1802--1813.

\bibitem{dean2008mapreduce}
J.~Dean and S.~Ghemawat, ``Mapreduce: simplified data processing on large clusters,'' \emph{Communications of the ACM}, vol.~51, no.~1, pp. 107--113, 2008.

\bibitem{zaharia2016apache}
M.~Zaharia, R.~S. Xin, P.~Wendell, T.~Das, M.~Armbrust, A.~Dave, X.~Meng, J.~Rosen, S.~Venkataraman, M.~J. Franklin \emph{et~al.}, ``Apache spark: a unified engine for big data processing,'' \emph{Communications of the ACM}, vol.~59, no.~11, pp. 56--65, 2016.

\bibitem{shen2020magnet}
M.~Shen, Y.~Zhou, and C.~Singh, ``Magnet: push-based shuffle service for large-scale data processing,'' \emph{Proceedings of the VLDB Endowment}, vol.~13, no.~12, pp. 3382--3395, 2020.

\bibitem{dageville2016snowflake}
B.~Dageville, T.~Cruanes, M.~Zukowski, V.~Antonov, A.~Avanes, J.~Bock, J.~Claybaugh, D.~Engovatov, M.~Hentschel, J.~Huang \emph{et~al.}, ``The snowflake elastic data warehouse,'' in \emph{Proceedings of the 2016 International Conference on Management of Data}, 2016, pp. 215--226.

\bibitem{armenatzoglou2022amazon}
N.~Armenatzoglou, S.~Basu, N.~Bhanoori, M.~Cai, N.~Chainani, K.~Chinta, V.~Govindaraju, T.~J. Green, M.~Gupta, S.~Hillig \emph{et~al.}, ``Amazon redshift re-invented,'' in \emph{Proceedings of the 2022 International Conference on Management of Data}, 2022, pp. 2205--2217.

\bibitem{leis2014morsel}
V.~Leis, P.~Boncz, A.~Kemper, and T.~Neumann, ``Morsel-driven parallelism: a numa-aware query evaluation framework for the many-core age,'' in \emph{Proceedings of the 2014 ACM SIGMOD international conference on Management of data}, 2014, pp. 743--754.

\bibitem{trino-ft}
``Trino fault tolerance,'' \url{https://github.com/trinodb/trino/wiki/Fault-Tolerant-Execution}.

\bibitem{kreps2011kafka}
J.~Kreps, N.~Narkhede, J.~Rao \emph{et~al.}, ``Kafka: A distributed messaging system for log processing,'' in \emph{Proceedings of the NetDB}, vol.~11, 2011, pp. 1--7.

\bibitem{carbone2015apache}
P.~Carbone, A.~Katsifodimos, S.~Ewen, V.~Markl, S.~Haridi, and K.~Tzoumas, ``Apache flink: Stream and batch processing in a single engine,'' \emph{Bulletin of the IEEE Computer Society Technical Committee on Data Engineering}, vol.~36, no.~4, 2015.

\bibitem{lin2016streamscope}
W.~Lin, Z.~Qian, J.~Xu, S.~Yang, J.~Zhou, and L.~Zhou, ``$\{$StreamScope$\}$: Continuous reliable distributed processing of big data streams,'' in \emph{13th USENIX Symposium on Networked Systems Design and Implementation (NSDI 16)}, 2016, pp. 439--453.

\bibitem{zaharia2012resilient}
M.~Zaharia, M.~Chowdhury, T.~Das, A.~Dave, J.~Ma, M.~McCauly, M.~J. Franklin, S.~Shenker, and I.~Stoica, ``Resilient distributed datasets: A $\{$Fault-Tolerant$\}$ abstraction for $\{$In-Memory$\}$ cluster computing,'' in \emph{9th USENIX Symposium on Networked Systems Design and Implementation (NSDI 12)}, 2012, pp. 15--28.

\bibitem{shaikhha2018push}
A.~Shaikhha, M.~Dashti, and C.~Koch, ``Push versus pull-based loop fusion in query engines,'' \emph{Journal of Functional Programming}, vol.~28, p. e10, 2018.

\bibitem{neumann2011efficiently}
T.~Neumann, ``Efficiently compiling efficient query plans for modern hardware,'' \emph{Proceedings of the VLDB Endowment}, vol.~4, no.~9, pp. 539--550, 2011.

\bibitem{harizopoulos2005qpipe}
S.~Harizopoulos, V.~Shkapenyuk, and A.~Ailamaki, ``Qpipe: A simultaneously pipelined relational query engine,'' in \emph{Proceedings of the 2005 ACM SIGMOD international conference on Management of data}, 2005, pp. 383--394.

\bibitem{boncz2005monetdb}
P.~A. Boncz, M.~Zukowski, and N.~Nes, ``Monetdb/x100: Hyper-pipelining query execution.'' in \emph{Cidr}, vol.~5, 2005, pp. 225--237.

\bibitem{murray2013naiad}
D.~G. Murray, F.~McSherry, R.~Isaacs, M.~Isard, P.~Barham, and M.~Abadi, ``Naiad: a timely dataflow system,'' in \emph{Proceedings of the Twenty-Fourth ACM Symposium on Operating Systems Principles}, 2013, pp. 439--455.

\bibitem{akidau2013millwheel}
T.~Akidau, A.~Balikov, K.~Bekiro{\u{g}}lu, S.~Chernyak, J.~Haberman, R.~Lax, S.~McVeety, D.~Mills, P.~Nordstrom, and S.~Whittle, ``Millwheel: Fault-tolerant stream processing at internet scale,'' \emph{Proceedings of the VLDB Endowment}, vol.~6, no.~11, pp. 1033--1044, 2013.

\bibitem{gog2021falkirk}
I.~Gog, M.~Isard, and M.~Abadi, ``Falkirk wheel: Rollback recovery for dataflow systems,'' in \emph{Proceedings of the ACM Symposium on Cloud Computing}, 2021, pp. 373--387.

\bibitem{moritz2018ray}
P.~Moritz, R.~Nishihara, S.~Wang, A.~Tumanov, R.~Liaw, E.~Liang, M.~Elibol, Z.~Yang, W.~Paul, M.~I. Jordan \emph{et~al.}, ``Ray: A distributed framework for emerging $\{$AI$\}$ applications,'' in \emph{13th USENIX Symposium on Operating Systems Design and Implementation (OSDI 18)}, 2018, pp. 561--577.

\bibitem{zhuang2023exoflow}
S.~Zhuang, S.~Wang, E.~Liang, Y.~Cheng, and I.~Stoica, ``$\{$ExoFlow$\}$: A universal workflow system for $\{$Exactly-Once$\}$$\{$DAGs$\}$,'' in \emph{17th USENIX Symposium on Operating Systems Design and Implementation (OSDI 23)}, 2023, pp. 269--286.

\bibitem{flink-ft}
``An overview of end-to-end exactly-once processing in apache flink,'' \url{https://flink.apache.org/features/2018/03/01/end-to-end-exactly-once-apache-flink.html}.

\bibitem{wang2019lineage}
S.~Wang, J.~Liagouris, R.~Nishihara, P.~Moritz, U.~Misra, A.~Tumanov, and I.~Stoica, ``Lineage stash: fault tolerance off the critical path,'' in \emph{Proceedings of the 27th ACM Symposium on Operating Systems Principles}, 2019, pp. 338--352.

\bibitem{wang2021ownership}
S.~Wang, E.~Liang, E.~Oakes, B.~Hindman, F.~S. Luan, A.~Cheng, and I.~Stoica, ``Ownership: A distributed futures system for $\{$Fine-Grained$\}$ tasks,'' in \emph{18th USENIX Symposium on Networked Systems Design and Implementation (NSDI 21)}, 2021, pp. 671--686.

\bibitem{vuppalapati2020building}
M.~Vuppalapati, J.~Miron, R.~Agarwal, D.~Truong, A.~Motivala, and T.~Cruanes, ``Building an elastic query engine on disaggregated storage,'' in \emph{17th USENIX Symposium on Networked Systems Design and Implementation (NSDI 20)}, 2020, pp. 449--462.

\bibitem{raasveldt2019duckdb}
M.~Raasveldt and H.~M{\"u}hleisen, ``Duckdb: an embeddable analytical database,'' in \emph{Proceedings of the 2019 International Conference on Management of Data}, 2019, pp. 1981--1984.

\bibitem{polars}
``Polars,'' \url{https://github.com/pola-rs/polars}.

\bibitem{flight}
``Arrow flight,'' \url{https://arrow.apache.org/blog/2019/10/13/introducing-arrow-flight/}.

\bibitem{burns2016borg}
B.~Burns, B.~Grant, D.~Oppenheimer, E.~Brewer, and J.~Wilkes, ``Borg, omega, and kubernetes,'' \emph{Communications of the ACM}, vol.~59, no.~5, pp. 50--57, 2016.

\bibitem{spark-fail}
``Spark enhancements for elasticity and resiliency on amazon emr,'' \url{https://aws.amazon.com/blogs/big-data/spark-enhancements-for-elasticity-and-resiliency-on-amazon-emr/}.

\bibitem{lamb2012vertica}
A.~Lamb, M.~Fuller, R.~Varadarajan, N.~Tran, B.~Vandier, L.~Doshi, and C.~Bear, ``The vertica analytic database: C-store 7 years later,'' \emph{arXiv preprint arXiv:1208.4173}, 2012.

\bibitem{snowflake-ft}
``Mythbusting snowflake pricing! all the cool stuff you get with 1 credit,'' \url{https://medium.com/snowflake/mythbusting-snowflake-pricing-all-the-cool-stuff-you-get-with-1-credit-f3daad217a98}.

\bibitem{dataaisummit_vectordatalakes}
\BIBentryALTinterwordspacing
``Vector data lakes,'' Data + AI Summit by Databricks, 2023, [Accessed: 29-November-2023]. [Online]. Available: \url{https://www.databricks.com/dataaisummit/session/vector-data-lakes/}
\BIBentrySTDinterwordspacing

\bibitem{zhou2021foundationdb}
J.~Zhou, M.~Xu, A.~Shraer, B.~Namasivayam, A.~Miller, E.~Tschannen, S.~Atherton, A.~J. Beamon, R.~Sears, J.~Leach \emph{et~al.}, ``Foundationdb: A distributed unbundled transactional key value store,'' in \emph{Proceedings of the 2021 International Conference on Management of Data}, 2021, pp. 2653--2666.

\bibitem{sivasubramanian2012amazon}
S.~Sivasubramanian, ``Amazon dynamodb: a seamlessly scalable non-relational database service,'' in \emph{Proceedings of the 2012 ACM SIGMOD International Conference on Management of Data}, 2012, pp. 729--730.

\end{thebibliography}
\end{document}